\documentclass[twocolumn, trackchanges]{aastex701}

\newcommand{\kms}{ km\,$s^{-1}$}

\begin{document}

\title{A Monopolar Jet from Protostar HOPS 10: Evidence for Asymmetric Magnetized Launching}

\author[orcid=0000-0002-2338-4583,sname='Dutta']{Somnath Dutta}
\affiliation{Academia Sinica Institute of Astronomy and Astrophysics, No.1, Sec. 4, Roosevelt Rd, Taipei 106319, Taiwan, R.O.C}
\email[show]{sdutta@asiaa.sinica.edu.tw}

\begin{abstract}
A fundamental challenge in star formation is understanding how a protostar accretes mass from its circumstellar disk while removing excess angular momentum. Protostellar jets are widely invoked as the primary channels for angular-momentum removal, yet the mechanism by which they are launched and extract angular momentum remains poorly constrained. Here, we report high-resolution ALMA Band~7 (345\,GHz) and Band~6 (230\,GHz) observations of $^{12}$CO\,(3--2), $^{12}$CO\,(2--1), and SiO\,(5--4) emission from the protostar HOPS\,10 (G209.55$-$19.68S2). The combined data trace both the entrained outflow and the collimated jet with excellent spatial and velocity resolution, revealing a uniquely monopolar protostellar jet—the clearest example reported to date. The system exhibits a distinctly unipolar high-velocity jet  (velocity offset, $\rm V_{HV,off} = V_{\rm observed} - V_{\rm systemic}$ = $+$44 to $+$66 km s$^{-1}$), unlike the predominantly bipolar morphology characteristic of most protostellar jets. While the low-velocity outflow (velocity offset, $\rm V_{LV,off} = V_{\rm observed} - V_{\rm systemic}$ = $-$20 to $+$30 km s$^{-1}$) is detected in both directions, the high-velocity jet appears only on one side, and this monopolarity is consistent across all tracers. Given the nearly edge-on geometry and low submillimeter extinction, comparable emission would normally be expected from both lobes. The shock tracer SiO emission confirms a genuine, highly collimated jet rather than cloud contamination, and no ambient structure is capable of obscuring a counterjet. We argue that intrinsically asymmetric mass loading along the disk’s magnetic field lines provides the most plausible explanation for the observed monopolarity.
\end{abstract}



\section{Introduction} 
A fundamental  challenge in star formation is understanding how a protostar continues to accrete mass despite the angular momentum that would otherwise centrifugally support the material and inhibit inward collapse. Protostellar jets are among the most striking manifestations of star formation, providing critical clues to the mass–accretion and angular–momentum transport processes that regulate early stellar evolution. By removing excess angular momentum from the innermost disk, jets help stabilize the accretion flow and enable smooth mass transport onto the protostar. Despite their ubiquity, protostellar jets exhibit substantial morphological diversity, ranging from well-collimated bipolar structures to highly asymmetric or even monopolar systems. The physical origin of these asymmetries—whether arising from anisotropic extinction, strong environmental density gradients, or intrinsically unequal mass loading along the disk magnetic field lines—remains an open question, and resolving these possibilities is essential for understanding how jet launching couples to accretion dynamics.

Protostellar outflows are generally understood to consist of two dynamically linked components: a fast ($>50~\mathrm{km\,s^{-1}}$), high-density, highly collimated jet launched from the innermost disk, and a slower ($\sim 1$--$30~\mathrm{km\,s^{-1}}$), lower-density, wide-angle disk wind originating from larger radii \citep[e.g.,][]{2007prpl.conf..245A,2014ApJ...783...29D,2014prpl.conf..451F,2015A&A...576A.109Y,2016ARA&A..54..491B,2020A&ARv..28....1L,2024AJ....167...72D,2025Univ...11..333D}.  These jet--wind components are typically traced in optical and infrared forbidden lines and in H$_2$ molecular transitions \citep[e.g.,][]{2016ARA&A..54..491B,2021NewAR..9301615R,2025A&A...699A.361V}, as well as in submillimeter molecular tracers such as CO, SiO, SO, and H$_2$CO that probe entrained and shock-excited gas \citep[e.g.,][]{2020A&ARv..28....1L,2024AJ....167...72D,2025Univ...11..333D}.

A key challenge in interpreting observed monopolar or asymmetric jets lies in disentangling intrinsic launching asymmetry from observational bias. Optical and near-infrared tracers are often affected by severe extinction, while single-band millimeter observations can be limited by excitation effects and optical depth uncertainties. Multi-band millimeter/submillimeter studies therefore provide a powerful, extinction-free means to probe the jet–outflow system across a range of excitation conditions and physical depths. In particular, simultaneous observations of multiple CO transitions and SiO lines enable a comprehensive comparison between entrained molecular gas and high-velocity shocks, offering a robust diagnostic of jet collimation, excitation, and energetics.

To address these challenges, we conducted dual-band ALMA observations (Bands~6 and~7) of the protostar \object{HOPS\,10} (also known as G209.55$-$19.68S2), a Class~0 source located at a distance $\sim$ 380 pc in Orion. \object{HOPS\,10} has previously reported properties of bolometric temperature $T_{\mathrm{bol}} = 48 \pm 11$~K, luminosity $L_{\mathrm{bol}} = 3.4 \pm 1.4~L_\odot$, and envelope mass $M_{\mathrm{env}} = 0.61 \pm 0.12~M_\odot$ \citep[][]{2020ApJ...890..130T, 2020ApJS..251...20D, 2024AJ....167...72D}. Based on the morphology of the outflow shell, \citet{2024AJ....167...72D} estimated an inclination angle of $i = 20^{+10}_{-5}$ degree (here, $i = 90$ degree is pole-on jet or face-on disk).

In this Letter, we confirm the asymmetry of the jet using multi-band submillimeter diagnostics. The dual-band observations allow a direct comparison of molecular gas morphology and kinematics across different excitation regimes, providing a robust framework to probe the origin of jet monopolarity and to test magneto-hydrodynamic (MHD) jet-launching models under asymmetric conditions. Systems in which jet asymmetry can be confirmed through multi-band millimeter data are rare, making such cases particularly valuable. We describe the observations in Section~\ref{sec:observations}, present the jet analysis and protostellar mass estimation in Section~\ref{sec:analyses_results}, and discuss the possible origins of the observed monopolarity along with its implications for current jet and wind launching models in Section~\ref{sec:discussion}.

\section{Observations}\label{sec:observations}
\subsection{Band 7}
The ALMA Band~7 observations analyzed in this study were obtained under project  \texttt{2015.1.00041.S  \citep[PI: John Tobin,][]{2020ApJ...890..130T}} conducted in 2016-2017. We utilized  the {$^{12}$CO\,(3 $-$ 2) transition at 345.79599~GHz, the $^{13}$CO\,(3 $-$ 2) transition at 330.58797~GHz, and the 0.87\,mm continuum. The corresponding spectral setups provided bandwidths of 937.5~MHz with a native channel spacing of 0.489~\kms{} for {$^{12}$CO\,(3 $-$ 2), and 234.375~MHz with a native channel spacing of 0.128~\kms{} for $^{13}$CO\,(3 $-$ 2). The raw visibilities were calibrated using the ALMA pipeline within \textsc{CASA}~4.7.2, and the final imaging was carried out in \textsc{CASA}~6.6.1. Phase-only self-calibration was applied to improve the image fidelity. The {$^{12}$CO\,(3 $-$ 2) cube was imaged using a 0\farcs35 \texttt{uvtaper} and a velocity resolution of 2~\kms{}, yielding a synthesized beam of $\sim$ 0\farcs42~$\times$~0\farcs39 and a channel rms of $\sim$ $2\times10^{-2}$~Jy~beam$^{-1}$. This choice of \texttt{uvtaper} and velocity binning provides a spatial resolution comparable to the available Band~6 archive data while improving sensitivity and adequately sampling the broad jet and outflow line widths. As a well-established tracer of protostellar jets and outflows, this transition is ideally suited for mapping high-velocity jet emission. In contrast, the $^{13}$CO\,(3 $-$ 2) cube---a well-known tracer of warm disk and envelope gas---was imaged at 0.44~\kms{} velocity resolution, resulting in a synthesized beam of $\sim$ 0\farcs14~$\times$~0\farcs13 and a per-channel sensitivity of $\sim$ $2.7\times10^{-2}$~Jy~beam$^{-1}$. The continuum image achieves a beam size of $\sim$ 0\farcs114~$\times$~0\farcs107 with an rms of $\sim$ $4.8\times10^{-4}$~Jy~beam$^{-1}$. All imaging was carried out with the \texttt{tclean} task, adopting a Briggs robust weighting parameter of $+$2.0 (natural weighting) for all data products.

\subsection{Band 6}
We utilized ALMA Band\,6 data from the ALMASOP program (Project ID: 2018.1.00302.S; PI: Tie Liu, \citealt{2020ApJS..251...20D}), calibrated and imaged using \texttt{CASA}~5.5. The observations were conducted in October--November~2018 in three array configurations—the 12\,m C43--5 (TM1), 12\,m C43--2 (TM2), and the 7\,m Atacama Compact Array (ACA). The correlator setup simultaneously covered multiple molecular transitions; in this work, we use only the $^{12}$CO\,(2 $-$ 1) (230.462\,GHz) and SiO\,(5 $-$ 4) (217.033\,GHz) lines, both are widely used tracers of molecular jets. Imaging was performed with the \texttt{tclean} task at a velocity resolution of 2\,km\,s$^{-1}$, adopting Briggs weighting (robust = $+$2.0, natural weighting)). The combination of the TM1, TM2, and ACA datasets resulted in synthesized beams of $\sim$ 0\farcs44\,$\times$\,0\farcs35 for $^{12}$CO\,(2 $-$ 1) and $\sim$ 0\farcs48\,$\times$\,0\farcs38 for SiO\,(5 $-$ 4), with corresponding per-channel sensitivities of $\sim$ 0.02--0.2\,mJy\,beam$^{-1}$. We also produced ACA-only maps for the $^{12}$CO\,(2--1) emission and the 1.3\,mm continuum using similar \texttt{tclean} parameters, yielding lower angular resolution (synthesized beam $\sim$ 8\farcs01\,$\times$\,4\farcs37).

\begin{figure*}
    \centering
    \includegraphics[width=0.8\linewidth]{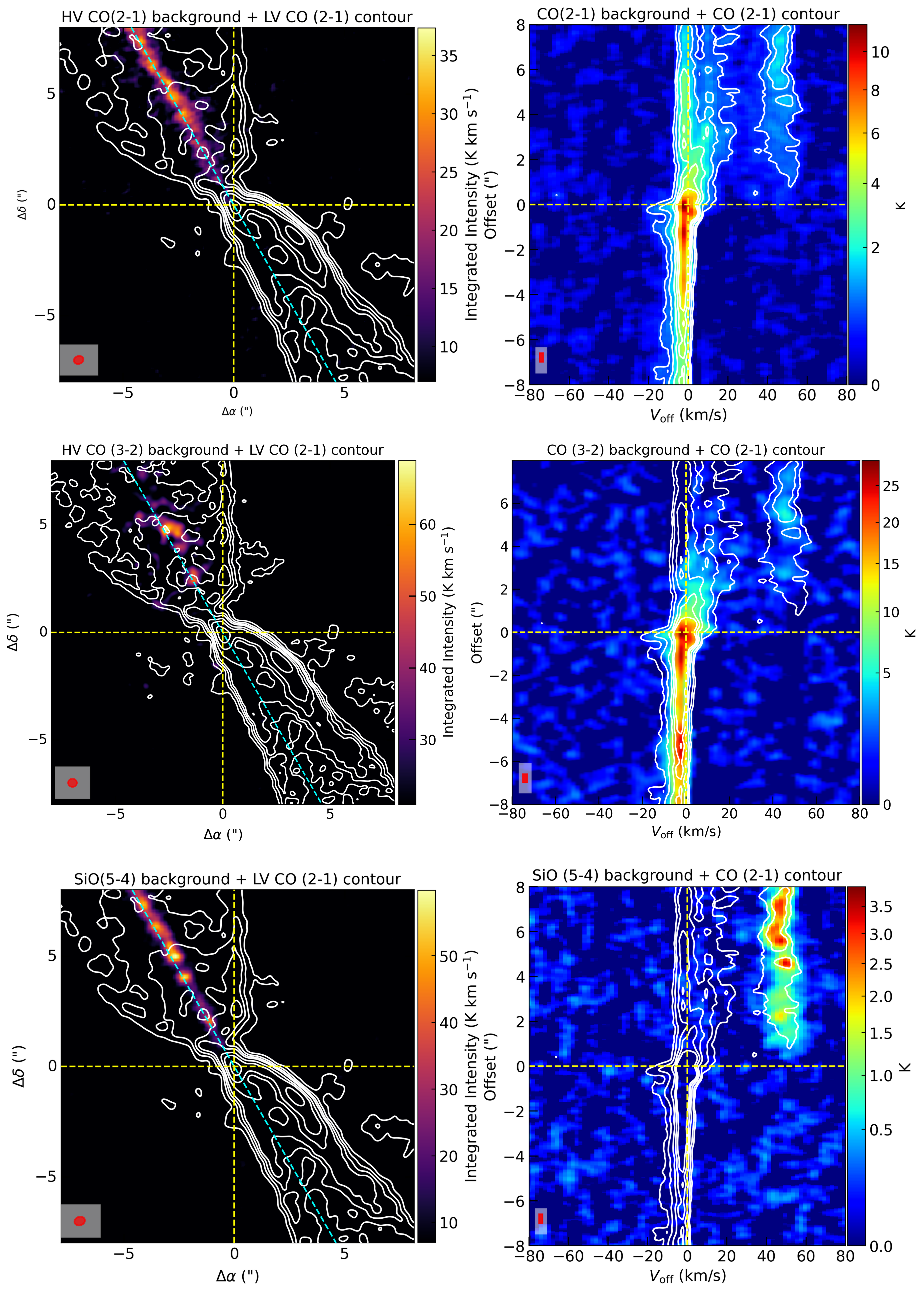}
    \caption{
     \textbf{Left column:} Integrated-intensity maps of $^{12}$\!CO(2--1), $^{12}$\!CO(3--2), and SiO(5--4), from top to bottom at $\sim$ 160 AU resolution. The background images show the integrated intensity of the high-velocity (HV) components ($V_{\rm HV, off} = V_{\rm obs} - V_{\rm sys} = +44$ to $+66~\mathrm{km\,s^{-1}}$), selected based on the position--velocity diagram in the right column. The sensitivities of the three maps are 3, 13, and 2.6~$\mathrm{K\,km\,s^{-1}}$, respectively. Contours trace the $^{12}$\!CO(2--1) emission integrated over the low-velocity (LV) range ($V_{\rm LV,off} = -20$ to $+30~\mathrm{km\,s^{-1}}$) above the $3\sigma$ level, and are drawn at $3n\sigma$ (where $n = 1, 2, 4, \ldots$ and $\sigma = 3~\mathrm{K\,km\,s^{-1}}$). The jet axis (PA = 30$\degr$ from North to East) is shown as a cyan dashed line, and the source position is at $(0,0)$. The synthesized beam is shown in the lower left of each panel. 
     \textbf{Right column:} Position--velocity (PV) diagrams extracted along the jet axis. The rms noise levels are $\sim$0.25~K for $^{12}$\!CO(2--1), $\sim$1.3~K for $^{12}$\!CO(3--2), and 0.24~K for SiO(5--4). $^{12}$\!CO(2--1) contours are overplotted in all three panels at 5, 10, 20, and 40\% of the peak emission. The source location is at $(0,0)$. The velocity and angular resolutions are indicated in red in the lower left corner of each panel.
    }
    \label{fig:combined_integrated_PV}
\end{figure*}

\section{Analysis and Results}\label{sec:analyses_results}

\subsection{Monopolar Collimated Jet and Bipolar Extended Winds}
Figure~\ref{fig:combined_integrated_PV} presents the $^{12}$CO(2--1), $^{12}$CO(3--2), and SiO(5--4) emission in the top, middle, and bottom panels, respectively. All maps are restored with nearly identical synthesized beams (corresponding to $\sim$160~AU resolution), ensuring a consistent comparison across the three transitions. The left column shows the high-velocity (HV) integrated intensity maps for the velocity range $\rm V_{HV,off} = V_{\rm obs} - V_{\rm sys} = +44$ to $+66~\mathrm{km\,s^{-1}}$, where the systemic velocity ($\rm V_{sys}\sim 8.29~km\,s^{-1}$) is determined from $^{13}$CO(3--2). The low-velocity (LV) $^{12}$CO(2--1) emission integrated over $\rm V_{LV,off} = -20$ to $\rm +30~km\,s^{-1}$ is overlaid as contours in all three panels for reference. The HV and LV intervals used for integration are derived directly from the corresponding position-velocity (PV) diagrams shown in the right column. Both the LV wind component and the HV jet component span nearly identical velocity ranges across all tracers, if detected. The mean velocity of HV jet and LV wind are $\rm V_{jet,off} = +54~(\pm 10) ~km\,s^{-1}$ and $\rm V_{wind,off} = \pm 14 ~(\pm 6) ~km\,s^{-1}$.  

The spectra extracted along the jet axis are provided in Appendix~\ref{sec:appendix_spectra}, and the low-resolution ($\sim$2000~AU) ACA channel maps are shown in Appendix~\ref{sec:appendix_ACACO_channelMap}. The apparent CO(3--2) emission feature is detected in the observed velocity range $V_{\rm obs} \approx -23 \pm 5~{\rm km~s^{-1}}$ in the Figure \ref{fig:appendix_spectra} (Appendix~\ref{sec:appendix_spectra}), corresponding to an offset velocity $V_{\rm HV, off} \approx -31~\pm 5~{\rm km~s^{-1}}$. It could therefore represent a blueshifted low signal-to-noise HV component. However, the emission is spatially isolated and not axially extended like the redshifted HV jet. Integration over this velocity range does not reveal emission above the $3\sigma$ level, and there is no corresponding detection in SiO(5--4) or CO(2--1) at any resolution, preventing a firm identification as a genuine blueshifted HV counterflow. 

The LV $^{12}$CO(3--2) emission is shown in Figure~\ref{fig:disk_jet_wind}, overlaid with LV $^{12}$CO(2--1) contours in black (as in Figure~\ref{fig:combined_integrated_PV}) and HV contours in red. The LV emission in both Figures~\ref{fig:disk_jet_wind} and ~\ref{fig:combined_integrated_PV} reveals two wide-angle bipolar lobes. The observed opening angles are $75.9^\circ \pm 3.8^\circ$ for the redshifted (north-east) lobe and $54^\circ \pm 4^\circ$ for the blueshifted (south-west) lobe, corresponding to deprojected opening angles of $72.5^\circ \pm 3.8^\circ$ and $52^\circ \pm 4^\circ$, respectively.  The velocity structure indicates that the wide-angle winds in both lobes are traced by this LV CO emission, whereas the HV component appears only on the redshifted side, exhibiting a series of knotty structures. The SiO emission traces only the HV component in the redshifted lobe (Figure \ref{fig:combined_integrated_PV}), further reinforcing this asymmetry. This redshifted jet in SiO emission shows an observed opening angle of $20^\circ \pm 6.5^\circ$ (or $18.8^\circ \pm 5.9^\circ$ when corrected for inclination). From the redshifted side alone, we measure a mean beam-deconvolved jet width of $\rm l_{wj} \sim 130 \pm 60~\mathrm{au}$. 

 We note that the asymmetry in the width of the LV outflow cavity, with the south--west lobe being more collimated than the north--east lobe, could be a consequence of the HV jet in the north--east lobe inducing additional lateral expansion to the LV outflow cavity. 
In addition, the position--velocity diagrams indicate a larger amount of gas at velocities close to the systemic velocity toward the south--west compared to the north--east, which may also contribute to the observed asymmetry in the low-velocity cavity by limiting the lateral expansion in the LV south--west outflow. However, more detailed modeling of the jet--outflow interaction would be required to confirm whether the jet can indeed drive such lateral expansion of the outflow cavity.

\begin{figure}
    \centering
    \includegraphics[width=1\linewidth]{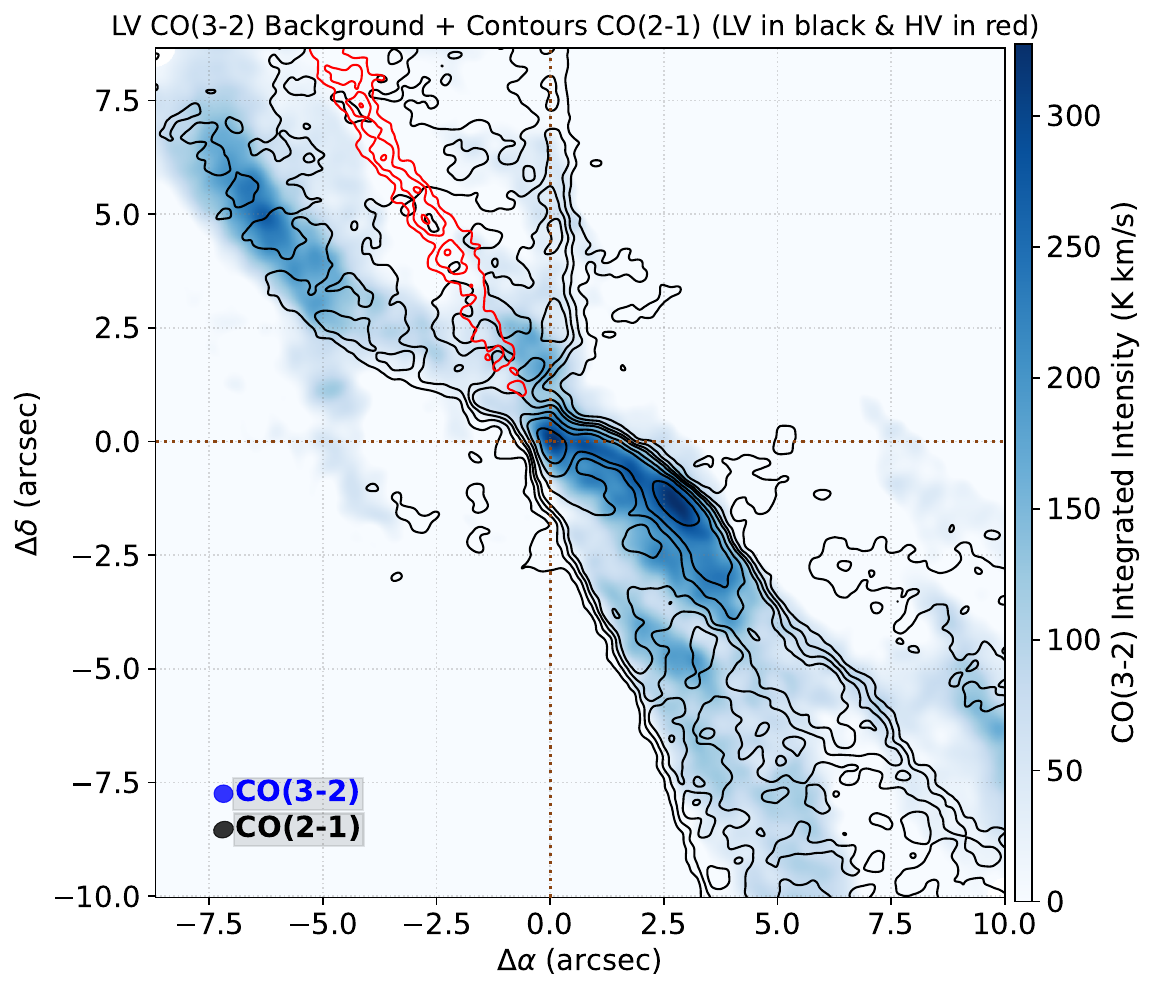}
    \caption{Background image showing the low-velocity (LV) $^{12}$CO\,(3--2) emission integrated over $\rm V_{LV,off}$ = $-20$ to $+30~\mathrm{km\,s^{-1}}$, having a sensitivity of $\sim 13~\mathrm{K\,km\,s^{-1}}$. Black contours indicate the $^{12}$CO\,(2--1) emission over the same LV range. The contours are drawn above the $3\sigma$ level at $3 n_1 \sigma$ (where $n_1 = 1, 2, 4, \ldots$ and $\sigma = 2.5~\mathrm{K\,km\,s^{-1}}$). Red contours trace the high-velocity (HV) $^{12}$CO\,(2--1) emission, $\rm V_{HV,off}$  = $+44$ to $+66~\mathrm{km\,s^{-1}}$, drawn at $3 n_2 \sigma$ (where $n_2 = 2, 3, 4, \ldots$ and $\sigma = 2.6~\mathrm{K\,km\,s^{-1}}$).}
    \label{fig:disk_jet_wind}
\end{figure}

We analyzed the CO~(3--2) and CO~(2--1) emission under the optically thin and local thermodynamic equilibrium (LTE) assumption to estimate the excitation conditions and opacity in the jet (Appendix~\ref{appendix_sec:opticaldepth}). After beam-matching and reprojecting the maps, we constructed a CO line--ratio image using only pixels above a $3\sigma$ threshold. The excitation temperature and initial column density, optical depth were computed using the radiative transfer solution. We find that the excitation temperature varies between 8.0 and 20\,K (mean  $\sim$ 9.0\,K), with optical depth $\tau_{\rm CO(2-1)}$ ranging from 0.03 to 6.9 (mean $\sim$ 0.54) and $\tau_{\rm CO(3-2)}$ ranging from 0.25 to 6.9 (mean $\sim$ 1.1). We note that these spatially compared estimates, derived from the line ratio at each location, are somewhat affected by proper motion, as the different tracers probe different regions and the observations are separated by $\sim$ 2.15 years. Taking jet velocity $\rm V_{jet} = V_{jet,off} / \sin \it{i}$, the CO~(2--1) emission may have shifted by $\sim0.19\arcsec$, approximately half of one beam resolution of these maps. We therefore measured the same parameters of nearest knots separated  by nearly half a beam width which turns out to be T$_{kin}$ $\sim$ 46\,K and $\tau_{\rm CO(2-1)}$ $\sim$ 0.04 and $\tau_{\rm CO(3-2)}$ $\sim$ 0.41. These values should therefore be interpreted with caution when inferring the physical conditions of the jet.
We also attempted pixel-wise non-LTE modeling using RADEX following \citet{2007A&A...468..627V}, but the fits were not reliable, likely due to the limited number of transitions. In all cases, the derived temperatures remained close to the initial guesses, and the resulting fit plots were not physically meaningful.

\subsection{Mass-loss Rates through Jets and Winds}
We derive the physical and kinematic properties of the jet primarily from the $^{12}$CO~(3--2) data. The CO~(3--2) transition is more easily excited in warm, shock-heated gas than CO~(2--1), which is more sensitive to the colder background material. In these observations, the $^{12}$CO~(3--2) data also offer superior velocity resolution relative to the $^{12}$CO~(2--1) data. This allows us to downsample the CO~(3--2) cube when necessary for a consistent comparison with the lower-resolution datasets, while retaining the advantages of the higher-frequency data.

\subsubsection{Wind Mass-Loss Rate Estimation} \label{sec:massloss_CO}

We derive the mass-loss rate for the low-velocity wind components separately from the high-velocity component, using the CO (3--2) line (shown as the background image in Figure~\ref{fig:disk_jet_wind}) and assuming optically thin and LTE conditions. The velocity integrated intensity in the low-velocity range  ($\rm V_{LV,off}= -20 ~to +30 ~ km\,s^{-1}$) is measured as the mean integrated intensity $\rm I_{\rm CO}$ (in K km s$^{-1}$) over the outflow area per pixel. Following \citet{2015A&A...576A.109Y,2015PASP..127..266M,2024AJ....167...72D}, the upper-state column density is calculated as
\begin{equation}\label{equ:Nu}
  \rm  N_u = \frac{8\pi k \nu^2}{h c^3 A_{ul}}\, I_{\rm CO},
\end{equation}
where $\rm \nu$ is the transition frequency and $\rm A_{ul}$ is the Einstein coefficient.  
The total CO column density is then obtained from
\begin{equation}\label{equ:NCO}
   \rm N_{\mathrm{CO}} = N_u \frac{Q(T_{\mathrm{ex}})}{g_u}\,
        \exp\!\left(\frac{E_u}{T_{\mathrm{ex}}}\right),
\end{equation}
using a partition function $Q(T_{\mathrm{ex}})$. Here,
$\rm g_u$ is the statistical weight of the upper level
($\rm g_u = 2J_u + 1$), and $E_u$ is the upper-level energy (in K).  We assume an excitation temperature for the outflow T$_{ex}$ = 50\,K \citep[e.g.][]{2024AJ....167...72D}. Assuming a \,CO/H$_2$ abundance ratio $\rm X_{CO} = 10^{-4}$, we infer the H$_2$ column density $\rm N_{H_2} = N_{CO}/X_{CO}$ and obtain the outflow mass as
\begin{equation}
    \rm M_{wind} = \mu_{H_2}\, m_{\mathrm{H}}\, N_{\mathrm{H}_2}\, A_{\mathrm{physical}},
\end{equation}
where $\rm A_{\mathrm{physical}}$ is the projected physical area, $\rm \mu_{H_2}$ = 2.8 is the molecular weight for hydrogen, and $m_{\mathrm{H}}$ is the mass of a hydrogen atom. 

The dynamical timescale is estimated as $\rm t_{dyn} = L/V_{wind}$, where $L$ is the projected outflow length in the field-of-view, and $V$ is the inclination-corrected velocity $\rm V_{wind} = V_{wind,off} / \sin \it{i}$. The wind mass-loss rate is then 
\begin{equation}
   \rm \dot{M}_{wind} = \frac{M_{wind}}{t_{\mathrm{dyn}}}.
\end{equation}

We have estimated the mass--loss rate of the red lobe (North--Eastern), $\rm \dot{M}_{\mathrm{wind,\,Red}} \sim (2.69 \pm  0.85)\times10^{-6}\,M_\odot\,\mathrm{yr^{-1}}$, and the blue lobe (South--Western), $\rm \dot{M}_{\mathrm{wind,\,Blue}} \sim (3.21 \pm 1.05)\times10^{-6}\,M_\odot\,\mathrm{yr^{-1}}$. These estimates are corrected for the inclination angle. We note that for a nearly edge--on system like HOPS\,10, the blue- and redshifted emission overlap in the velocity dimensions, as seen in the PV diagrams in Figure~\ref{fig:combined_integrated_PV}. This overlap makes it challenging to clearly separate the velocity space in the low-velocity channels close to the systemic velocity. Therefore, in our calculation we consider the full velocity range, including all overlapping channels.  Although we do not detect any large--scale molecular cloud emission that would significantly contaminate the outflow mass, we assume that any 
minimal cloud contribution would be comparable for both lobes. Our results should be regarded as upper limits. The uncertainties in  these estimates are dominated by the error in the inclination angle, velocities and measured fluxes.

\subsubsection{Jet Mass-Loss Rate}
To estimate the mass loss rates through the HV jet,  we have taken a slightly different approach than wind-mass loss rate estimation. We model the jet as a cylindrical flow of gas primarily composed of molecular hydrogen, with a mean column density $\rm N(H_2)$ and a constant jet velocity $\rm V_{jet} = V_{jet, off}/\sin \it{i}$. Under standard assumptions of optically thin and LTE conditions, the CO column density is measured from the total integrated intensity ($\rm I_{\rm CO}$ in K km s$^{-1}$) in the HV range ($\rm V_{HV,off}= +44 ~to ~+66 ~ km\,s^{-1}$) following equations \ref{equ:Nu} and \ref{equ:NCO}. We adopt same $X_{\rm CO} = 10^{-4}$ and slightly higher temperature $T_{\rm ex} = 150$~K for the dense jet than the wind-components, consistent with previous jet studies \citep[e.g.,][]{2024AJ....167...72D}. The mass-loss rate through the jet can be approximated by the following expression:

\begin{equation}
  \rm  \dot{M_{jet}} = \mu_{H_2} m_{H} N(H_2) V_{jet} l_{wj}
\end{equation}

where $\mu_{H_2} = 2.8$ is the mean molecular weight per hydrogen molecule, $m_{\mathrm{H}}$ is the mass of a hydrogen atom, and $\rm l_{wj}$ is the width of the jet perpendicular to the flow direction, as estimated in the above section.  
We estimate a mass–loss rate for the red–shifted jet of $\rm \dot{M}_{jet,{\rm Red}} \sim 0.66^{+0.67}_{-0.45} \times 10^{-6}\, M_\odot\,{\rm yr^{-1}}$. The error bars in the estimation are propagated from inclination angle, velocity uncertainties, flux estimation, and jet width estimation. 

\begin{figure*}
    \centering
    \includegraphics[width=1\linewidth]{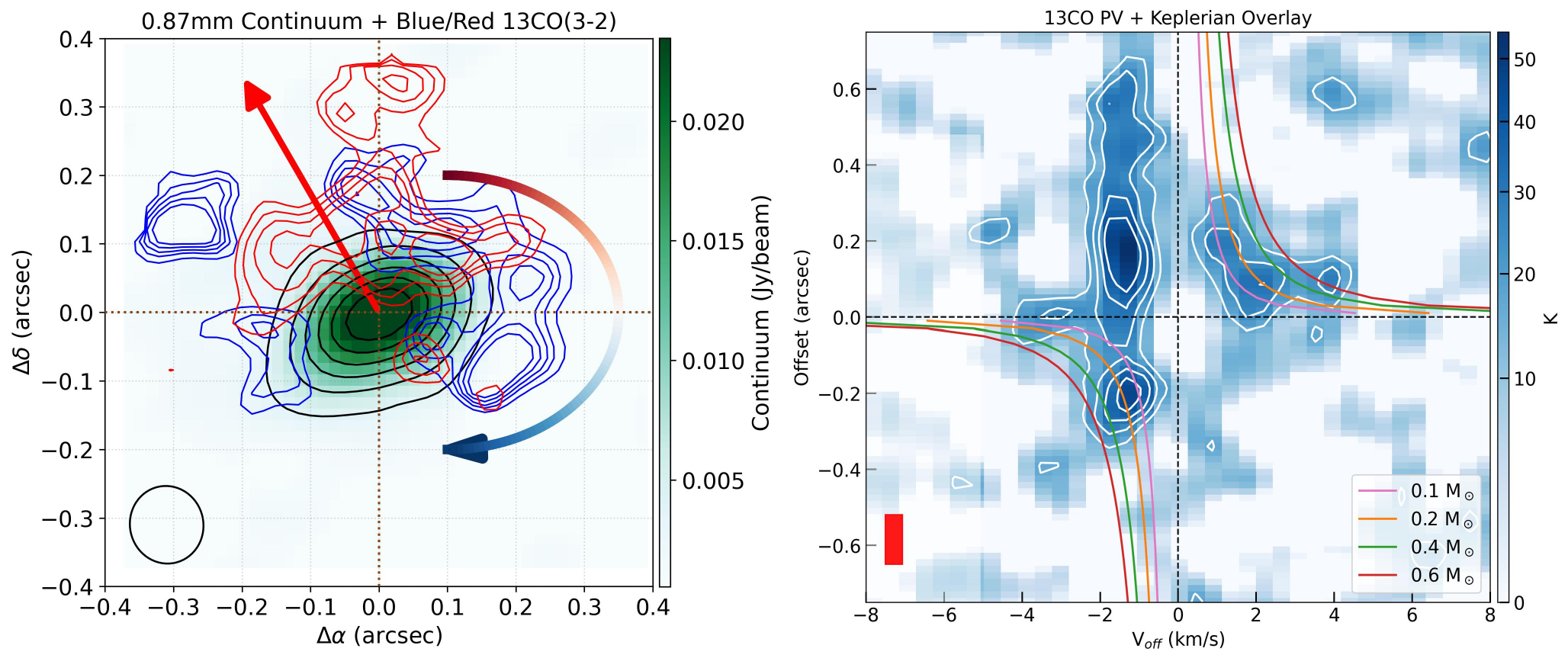}
    \caption{(Left panel) Integrated blue- and red-shifted \({}^{13}\)CO emission contours overplotted on the 0.87\,mm continuum emission. Black contours show the continuum at levels of \(3 n_1 \sigma\), where \(n_1 = 1,2,3,4,\ldots\) and sensitivity \(\sigma = 1.33\times10^{-4}~\mathrm{Jy\,beam^{-1}}\). Blue contours are integrated over \(V_{\rm obs} = 5.84\) to \(7.6~\mathrm{km\,s^{-1}}\) and drawn at \((70, 78, 83, 88, 93)\sigma\) with  sensitivity \(\sigma = 10~\mathrm{K\,km\,s^{-1}}\). Red contours are integrated over \(V_{\rm obs} = 8.92\) to \(10.68~\mathrm{km\,s^{-1}}\) and drawn at \((40, 48, 56, 64)\sigma\) with  sensitivity \(\sigma = 6~\mathrm{K\,km\,s^{-1}}\). A tentative sense of rotation is indicated by the curved arrow from red- to blue-shifted emission. The synthesized beam is shown in the lower left corner, and the red arrow marks the direction of the redshifted jet axis.
   (Right panel) Position--velocity (PV) diagram of the \({}^{13}\)CO emission along the disk major axis. White contours are drawn at 20, 30, 40, and 50\% of the peak emission. Keplerian rotation curves for several representative central masses are overplotted. Velocities are shown as offsets from the systemic velocity, \(V_{\rm off} = V_{\rm obs} - v_{\rm sys}\). The spatial and velocity resolutions are indicated in red in the lower left corner.    
    }
    \label{fig:13CO_diks_combined}
\end{figure*}


\subsection{Rotating Envelope and disk}\label{sec:pv_kepler_fit}

Figure~\ref{fig:13CO_diks_combined} (left panel) shows the 0.87\,mm continuum emission. From the deconvolved 0.87\,mm continuum size, we estimate the disk inclination assuming an intrinsically circular disk, $i_d = \cos^{-1}\!\left(\frac{b}{a}\right)$ 
where \(a\) and \(b\) are the major and minor axes, respectively. This yields a disk inclination of \(i_d \approx 60^\circ\), where \(i_d = 0^\circ\) corresponds to a face-on disk and \(i_d = 90^\circ\) to an edge-on disk.

This value is broadly consistent with the inclination inferred from the outflow shell geometry by \citet{2024AJ....167...72D}, who derived \(i = 20^{+10}_{-5}{}^\circ\) using a convention in which \(i = 0^\circ\) corresponds to an edge-on disk. Under this definition, the two inclinations are related by $\rm i_d = 90^\circ - i$ . Given that Class~0 disks may be geometrically thick, which can bias inclination estimates derived from continuum fitting, we adopt \(i_d = 70^\circ\) (corresponding to \(i = 20^\circ\)) in the subsequent kinematic analysis.

Figure~\ref{fig:13CO_diks_combined} (left panel) shows the inner-envelope emission at $\sim$ 45 AU resolution traced by the $\rm {}^{13}CO(3-2)$ line, with red- and blue-shifted components indicated relative to the systemic velocity and overplotted on the 0.87\,mm continuum emission.  Because the system is viewed nearly edge-on, the low-velocity red- and blue-shifted emission cannot be cleanly separated spatially and therefore overlaps in the integrated emission. Nevertheless, we indicate a tentative sense of rotation in the inner envelope with a rotating error. Emission from the disk itself around the continuum peak is not detected above  the \(3\sigma\) level possibly because the emission becomes optically thick.

The right panel of Figure~\ref{fig:13CO_diks_combined} presents the corresponding PV diagram of the $\rm {}^{13}CO(3-2)$ emission. To construct the PV diagram, we extracted a slice along the disk major axis and plotted the smoothed intensity distribution together with contour levels to highlight the emission structure. For a Keplerian disk, the radial (infall) velocity is negligible, \(V_r(r) \approx 0\), and the observed velocities in the PV diagram therefore trace the rotational component \(V_\phi(r)\). The Keplerian rotation profile is given by \citep[e.g.,][]{2016ApJ...826..213L}:
\begin{equation}
V_\phi(r) = \pm \sqrt{\frac{G\,M}{r}}\,\sin i_d ,
\end{equation}
where \(r\) is the deprojected radius, \(i_d\) is the disk inclination, and \(M = M_* + M_d\) is the enclosed mass, with \(M_*\) and \(M_d\) denoting the stellar and disk masses, respectively. Several representative Keplerian rotation curves spanning a range of central masses were overlaid on the PV diagram for visual comparison.  During the fitting procedure, we excluded inner radii affected by beam smearing by requiring a minimum spatial offset relative to the synthesized beam, in order to avoid biased peak velocities. The rotating inner envelope traced by \({}^{13}\)CO yields an estimated combined central protostar+disk mass in the range \(0.1\mbox{--}0.4~M_\odot\); we therefore adopt a fiducial mass of \(0.25 \pm 0.15~M_\odot\). This estimate is subject to uncertainties arising from the assumed distance and disk inclination, as well as additional systematic effects such as projection, non-Keplerian motions (e.g., outflow contamination, infall, or turbulence), and the limited spatial and spectral resolution of the observations. The inferred protostellar mass is comparable to those of other well-studied low-mass Class~0 protostellar systems, such as B\,335, HH\,211, and HH\,212 \citep{2017ApJ...834..178Y,2020A&ARv..28....1L}.

\section{Discussion}\label{sec:discussion}

\subsection{Detection of a Monopolar High-velocity Jet}\label{sec:mono_detection}

The HV jet emission is detected exclusively in the north-eastern lobe across all three tracers—$^{12}$CO(3--2), $^{12}$CO(2--1), and SiO(5--4)—yielding a clearly monopolar morphology in both ALMA Band~6 and Band~7 observations. Although the CO transitions probe slightly different spatial and excitation conditions, their HV emission spans the same velocity range, $+44$ to $+64~\mathrm{km,s^{-1}}$, as traced by SiO. No blueshifted HV emission is detected in any tracer, even upon detailed inspection of individual channel maps.

In contrast, the LV component, spanning $-20$ to $+30~\mathrm{km\,s^{-1}}$, is clearly detected on both sides of the source in CO transitions, forming a well-defined bipolar shell. Thus, while the wide-angle molecular outflow remains bipolar, the collimated HV jet is present only in the redshifted lobe.


 If we consider a jet knot composed of an ellipsoidal shock structure consisting of multiple layers—including a forward shock, a radiatively cooled post-shock layer, and a backward (reverse) shock—as described in jet–shock models \citep[e.g.,][]{2001ApJ...557..429L,2024AJ....167...72D}, an edge-on geometry can indeed facilitate the spatial separation of different radiative cooling zones at near-infrared wavelengths (e.g., with HST or JWST). However, the situation is fundamentally different for submillimeter molecular line observations. In submillimeter CO or SiO emission, the innermost radiatively cooled layers of a high-velocity jet knot may not be fully resolved even in a nearly edge-on system. This effect arises from a combination of optical depth, limited angular resolution, and line-of-sight projection, which can cause multiple shocked layers to overlap both spatially and kinematically in molecular tracers. Emission from shock layers located farther from the observer may also be partially obscured by intervening radiatively cooled material due to optical depth.

In the idealized case of a perfectly edge-on view, all shock layers could in principle be observed, whereas in a pole-on configuration the forward and backward shocks would appear predominantly blueshifted and redshifted, respectively. For intermediate inclinations, a mixture of these components is expected to contribute to the molecular emission. However, for HOPS 10, which is viewed at a nearly edge-on or modestly inclined geometry, these effects alone cannot account for the complete absence of the blueshifted high-velocity jet component, implying that the observed monopolarity is a robust observational result. While self-absorption may affect low-velocity CO emission near the systemic velocity, it is unlikely to influence the high-velocity regime discussed here.

\subsection{Is the Jet Intrinsically Monopolar?}\label{sec:mono_question}

Despite the one-sided appearance of the HV jet, the source drives a bipolar wide-angle molecular outflow. An important additional constraint is provided by the mass-loss rates inferred for the two lobes. In the redshifted lobe, the combined mass-loss rate due to the wind and jet is
$\dot{M}_{\mathrm{wind,\,red}} \sim (2.69 \pm 0.85)\times10^{-6}\,M_\odot\,\mathrm{yr^{-1}}$ and
$\dot{M}_{\mathrm{jet,\,red}} \sim 0.66^{+0.67}_{-0.45}\times10^{-6}\,M_\odot\,\mathrm{yr^{-1}}$,
respectively. This total is comparable, within uncertainties, to the mass-loss rate of the blue lobe, which is measured from the wide-angle wind alone,
$\dot{M}_{\mathrm{wind,\,blue}} \sim (3.21 \pm 1.05)\times10^{-6}\,M_\odot\,\mathrm{yr^{-1}}$.

Although the uncertainties are substantial, the near equality of the total mass-loss rates on the two sides suggests that the jet/outflow system is globally symmetric in mass-loss. This raises a fundamental question: is the jet intrinsically monopolar, or does the apparent absence of the counter-jet arise from projection effects, opacity, environmental asymmetry, or other observational biases that selectively suppress the blueshifted HV emission?

In the following sections, we critically examine these possibilities and assess whether they can account for the observed monopolarity, before considering intrinsic asymmetries in the jet-launching mechanism as the most plausible explanation.

\subsection{Alternative Explanations for the Apparent Monopolarity}\label{sec:mono_alternatives}

Before invoking an intrinsically asymmetric jet-launching mechanism, we examine whether the observed monopolarity of the high-velocity jet could arise from projection effects, opacity, environmental asymmetry, or time-dependent launching.

\paragraph{Projection effects and Doppler boosting.}
A favorable inclination can, in principle, cause only one side of a shock layer to be visible, producing an apparently monopolar jet \citep[e.g.,][]{2024AJ....167...72D}. However, because the system is viewed with a nearly edge-on inclination, both jet lobes should experience similar projection effects, with only minor brightness asymmetries expected due to the inclination not being perfectly edge-on. In the present source, the redshifted HV emission is detected in multiple tracers with distinct excitation conditions, appearing scattered in CO\,(3--2) but smooth and well collimated in SiO\,(5--4) and CO\,(2--1). This systematic, tracer-dependent behavior is inconsistent with a simple projection effect as the primary cause of the observed monopolarity.

While relativistic Doppler boosting can strongly enhance one-sided emission in AGN jets \citep[e.g., M\,87; ][]{2007ApJ...658..232C,2007ApJ...668L..27K}, the low deprojected velocity of this protostellar jet ($\sim 158~\mathrm{km~s^{-1}}$ for HOPS\,10) and its low stellar mass ($\sim 0.25 \pm 0.15~M_\odot$) make Doppler effects negligible. Therefore, Doppler boosting cannot explain the systematic absence of the blueshifted HV emission across all tracers in this case.

\paragraph{Optical depth and evolutionary effects.}
A non-detection of one jet lobe in a single transition could, in principle, be attributed to optical depth effects. However, in this case the redshifted high-velocity (HV) jet is robustly detected in both ALMA Band~6 and Band~7. The optical depths of all relevant transitions are generally modest ($\tau < 1$) throughout most of the redshifted jet, with occasional higher values at some scattered points, possibly caused by the shift of emission due to proper motion. Furthermore, in the redshifted HV jet, the estimated optical-depth hierarchy,
\[
\tau\bigl(^{12}\mathrm{CO}\,(3\!-\!2)\bigr) 
\;>\;
\tau\bigl(\mathrm{SiO}\,(5\!-\!4)\bigr)
\;>\;
\tau\bigl(^{12}\mathrm{CO}\,(2\!-\!1)\bigr),
\]
makes it highly unlikely that an entire HV counter-jet would be missed simultaneously across tracers with different optical depths.

Jet emission is expected to become increasingly ionized as protostars evolve, so most evolved protostars exhibit optical and infrared jets but no submillimeter molecular jet detection \citep[][]{2016ARA&A..54..491B,2021NewAR..9301615R}. It is also improbable that the jet would be ionized in only one direction, particularly given the low mass of the central protostar and the young, molecular nature of the jet. Although future low-frequency free--free observations or high-sensitivity infrared imaging with \textit{JWST} could provide additional constraints, optical depth or evolutionary effects are unlikely to account for the observed monopolarity.

Self-absorption in CO transitions may affect the low-velocity emission near the systemic velocity, where foreground envelope material can partially absorb the outflow emission. However, such effects are negligible at high velocities, where the jet emission is kinematically well separated from the ambient cloud. The absence of a blueshifted high-velocity jet therefore cannot be attributed to CO self-absorption. In contrast, SiO\,(5--4) traces mostly shocked gas, especially at high velocities, making it insensitive to foreground absorption. The lack of a blueshifted high-velocity SiO component thus strongly supports the intrinsic nature of the observed monopolarity.

\paragraph{Environmental asymmetry and extinction.}
An intrinsically bipolar jet may appear one-sided if the counter-jet propagates into a denser or more extinguished region of the surrounding envelope, leading to enhanced cooling, deceleration, or obscuration. Such environmental asymmetries could suppress counter-jets in some embedded systems. In addition, jets that are intrinsically bipolar may appear absent or strongly attenuated at optical or infrared wavelengths due to line-of-sight differential extinction by circumstellar dust, even when the underlying jet remains physically present \citep{2001AJ....122..432R,2007prpl.conf..245A,2025ApJ...991...45D,2025Univ...11..333D}. 

In the present case, the low-velocity molecular outflow is clearly bipolar, and the lobe lacking the HV jet in fact shows a higher mass-loss rate in the wide-angle wind. Moreover, $^{13}$CO\,(2--1) data at $\sim$ 45~AU resolution, $^{12}$CO\,(2--1) at $\sim$150~AU, and larger-scale observations probing $\sim$ 2000~AU scales (Appendix~\ref{sec:appendix_ACACO_channelMap}), together with published low-resolution continuum data \citep{2020ApJS..251...20D,2024AJ....167...72D}, reveal no dense material or significant asymmetric envelope structure capable of obscuring or quenching a jet. At millimeter wavelengths, extinction effects are negligible, and at all spatial scales we find no evidence for a blueshifted HV jet or for large-scale cloud structures that could suppress faint jet emission.

\paragraph{Episodic or time-variable jet launching.}
Time-dependent accretion or magnetic reconnection events can temporarily favor ejection in one hemisphere, producing a transiently asymmetric jet \citep[e.g.,][]{2015ApJ...814..113S}. If such an event had occurred recently, one would expect signatures of an accretion burst or a localized, isolated jet feature. In this source, however, the bolometric luminosity is low ($\rm3.2~\pm~1.4~L_\odot$), and the jet morphology consists of a continuous chain of knots rather than a single recent ejection. These characteristics argue against episodic launching as the primary origin of the observed monopolarity, although it cannot be entirely excluded.

\medskip
In summary, projection effects, optical depth or evolutionary effects, environmental asymmetry, and episodic launching all fail to account for the complete absence of a blueshifted high-velocity jet. We therefore conclude that the observed monopolarity most likely reflects an intrinsic asymmetry in the jet-launching process itself, which we discuss in the context of MHD wind models in the following section.

\subsection{Origin of the Monopolar Jet: Magnetic Asymmetry in Jet Launching}
\label{sec:mono_magnetic}

Protostellar jets are widely understood as magneto-centrifugally launched from rotating, magnetized disks, in which magnetic fields extract angular momentum from the disk and accelerate material along open field lines \citep{1982MNRAS.199..883B,1986ApJ...301..571P,1994ApJ...429..781S,1997A&A...319..340F}. The rotation of the disk twists the poloidal magnetic field, generating the toroidal component required for jet collimation. Our $^{13}$CO kinematic analysis (section~\ref{sec:pv_kepler_fit}) reveals ordered rotation in the inner envelope and/or disk, indicating the presence of a coherent angular-momentum reservoir capable of driving a magneto-centrifugal wind and jet. This provides direct observational support for a disk--jet connection in this system. Consistently, transverse velocity gradients observed in other protostellar jets have been interpreted as signatures of jet rotation, further supporting magneto-centrifugal launching from rotating disks \citep{2002ApJ...576..222B,2007ApJ...663..350C,2017NatAs...1E.152L}.

In the framework of MHD winds, protostellar outflows are often described as a two-component system, consisting of a wide-angle wind launched over several astronomical units from the disk \citep[disk-wind;][]{2007prpl.conf..277P} and a highly collimated, high-velocity jet originating from the innermost disk or stellar magnetosphere \citep[X-wind;][]{1994ApJ...429..781S}. In practice, both disk winds and X-winds are capable of producing collimated jets and broader molecular outflows \citep{2020ApJ...905..116S}, suggesting that the distinction between jets and winds reflects differences in launching radius rather than fundamentally different mechanisms. Crucially, the innermost launching region ($r \lesssim 0.1$~AU) is particularly sensitive to small hemispheric  variations in magnetic topology and mass loading \citep[e.g.,][]{2013ApJ...774...12F}.

Magneto-centrifugal jet launching depends critically on the strength, polarity, and inclination of the poloidal magnetic field threading the disk. Numerical simulations show that modest hemispheric differences—arising from non-axisymmetric accretion, misalignment between the stellar dipole and disk rotation axis, or MRI-driven turbulence—can strongly modify the mass loading onto magnetic field lines \citep{2006A&A...453..785F,2012A&A...545A..53M,2013ApJ...774...12F}. Because the jet mass flux depends sensitively on local disk surface density, ionization fraction, and field-line inclination, even small perturbations can produce order-of-magnitude differences between the two hemispheres, enabling efficient jet formation on one side while suppressing or quenching the counter-jet \citep{2010ApJ...722.1325S,2012A&A...545A..53M,2013ApJ...774...12F,2013A&A...550A..61L,2013A&A...550A..99Z,2015MNRAS.450..481D,2025ApJ...988..107T}. This unequal mass loading provides a particularly robust mechanism for generating monopolar jet.  In this scenario, a bright, fast jet emerges from only one side of the disk, while the larger-scale disk wind remains largely unaffected, naturally explaining the coexistence of a monopolar high-velocity jet with a symmetric wide-angle outflow.

This framework naturally explains the coexistence of a symmetric wide-angle CO outflow with a monopolar high-velocity jet in the HOPS\,10 protostar. The combined disk wind and jet wind preserve global bipolar symmetry in mass and momentum, while the innermost launching region remains susceptible to strong hemispheric asymmetries in magnetic topology and mass loading. Consequently, the system can sustain a persistently one-sided high-velocity jet without violating the underlying bipolar nature of magneto-centrifugal launching. Such behavior suggests that monopolar jets may represent an extreme but physically natural outcome of MHD jet-launching processes.

\section{Summary and Conclusion}

We have presented a detailed analysis of a rare monopolar protostellar jet, constituting the clearest example reported to date.

\begin{itemize}

\item 
High-velocity emission in $^{12}$CO(3--2), $^{12}$CO(2--1), and SiO(5--4) is detected exclusively on the redshifted side, demonstrating that the jet is intrinsically monopolar across multiple tracers and excitation conditions.

\item
In contrast, the low-velocity molecular outflow traced by $^{12}$CO(3--2) and $^{12}$CO(2--1) is detected in both lobes, indicating that the monopolarity is confined to the high-velocity jet component rather than the larger-scale low-velocity outflow.

\item 
The $^{13}$CO(2--1) data reveal a rotating inner envelope. Dynamical modeling yields a central protostellar mass of $\rm \sim0.25 \pm 0.15\,M_\odot$, indicating that even a very young, low-mass source can launch a highly collimated, one-sided jet.

\item  
Among the mechanisms examined, asymmetric mass loading or a hemispheric difference in magnetic topology provides the most compelling explanation for sustaining such monopolarity.

\end{itemize}
Future high angular resolution and high sensitivity polarimetric observations and advanced MHD modeling will be essential for constraining the magnetic topology and mass-loading asymmetry that may give rise to such a persistent one-sided jet.

\begin{acknowledgments} I am thankful to the anonymous reviewer for insightful comments that helped improve the manuscript. 
I would like to thank Dr. Chin-Fei Lee, Dr. Sheng-Yuan Liu, Dr. Tie Liu, and other members of the ALMASOP team for their support with ALMA data access, computational resources, and for valuable discussions regarding the data and scientific interpretation at various stages of my  research  over the past few years. This paper makes use of the following ALMA data:  ADS/JAO.ALMA$\#$2015.1.00041.S and $\#$2018.1.00302.S. ALMA is a partnership of ESO (representing its member states), NSF (USA) and NINS (Japan), together with NRC (Canada), NSC and ASIAA (Taiwan), and KASI (Republic of Korea), in cooperation with the Republic of Chile. The Joint ALMA Observatory is operated by ESO, AUI/NRAO and NAOJ.
\end{acknowledgments}

\begin{contribution}
SD conceptualized and developed the scientific idea, retrieved and analyzed the ALMA archival data, performed the raw data reduction, carried out the full scientific analysis, and led the writing of the manuscript. 
\end{contribution}

%
\vspace{5mm}
\facility{ALMA}\\

\software{Astropy \citep[][]{2013A&A...558A..33A,2018AJ....156..123A}, APLpy \citep[][]{2012ascl.soft08017R}, Matplotlib \citep[][]{2007CSE.....9...90H}, scipy \citep[][]{2020NatMe..17..261V}, CASA \citep[][]{2007ASPC..376..127M}, CARTA \citep[][]{2021ascl.soft03031C}.}\\

\appendix

\section{Spectra of HOPS10}\label{sec:appendix_spectra}
Figure \ref{fig:appendix_spectra} shows the spectra extracted from a $13\arcsec \times 1\arcsec$ rectangular region centered on the continuum peak and oriented along the jet axis (position angle $30^\circ$ east of north). A vertical line marks the systemic velocity, $\rm V_{sys}=8.29~{ km~s^{-1}}$. All spectra are smoothed using a Gaussian kernel with $\sigma = 2$ (implemented with \texttt{scipy.gaussian\_filter1d}) to improve the signal-to-noise ratio. 

The jet emission is clearly detected in the redshifted lobe, with prominent high-velocity components at $\rm V_{HV,off} = +44$ to $+66~{\rm km~s^{-1}}$. In contrast, no comparably peak is identified in the blueshifted velocity range. 
 The apparent CO(3--2) emission feature is detected in the observed velocity range $V_{\rm obs} \approx -23 \pm 5~{\rm km~s^{-1}}$, corresponding to an offset velocity $V_{\rm HV,off} = V_{\rm obs} - V_{\rm sys} \approx -31 \pm 5~{\rm km~s^{-1}}$. In this sense, it could represent a blueshifted HV component with a relatively low signal-to-noise ratio. However, a close look to the individual channels suggest that this emission arises from a spatially isolated location rather than from an axially extended structure as seen in the redshifted HV jet. It may instead correspond to an isolated emission in the jet or a localized velocity feature unrelated to a sustained counter-jet. When we integrated the emission solely within this velocity range, we were unable to identify emission above the $3\sigma$ level. Moreover, there
is no corresponding detection in either SiO(5--4) or CO(2--1) at any angular resolution, which prevents us from confidently identifying this feature as a genuine HV counterflow. Future observations with higher sensitivity may help to clarify this scenario. Importantly, even if the presence of such a weak blueshifted HV component were to be confirmed in an isolated location, the strong asymmetry between the redshifted and blueshifted jets would still persist. Therefore, our main conclusion regarding the intrinsic asymmetry of the jet remains unchanged.

We note that the jet emission appears relatively faint in these spectra because the extraction region averages over a large area; however, when inspecting individual channel maps along the jet axis, the jet features are several orders of magnitude stronger relative to the local sensitivity.

\setcounter{figure}{0} 
\renewcommand{\thefigure}{A\arabic{figure}} 
\begin{figure}[ht!]
\centering
\includegraphics[width=0.8\linewidth]{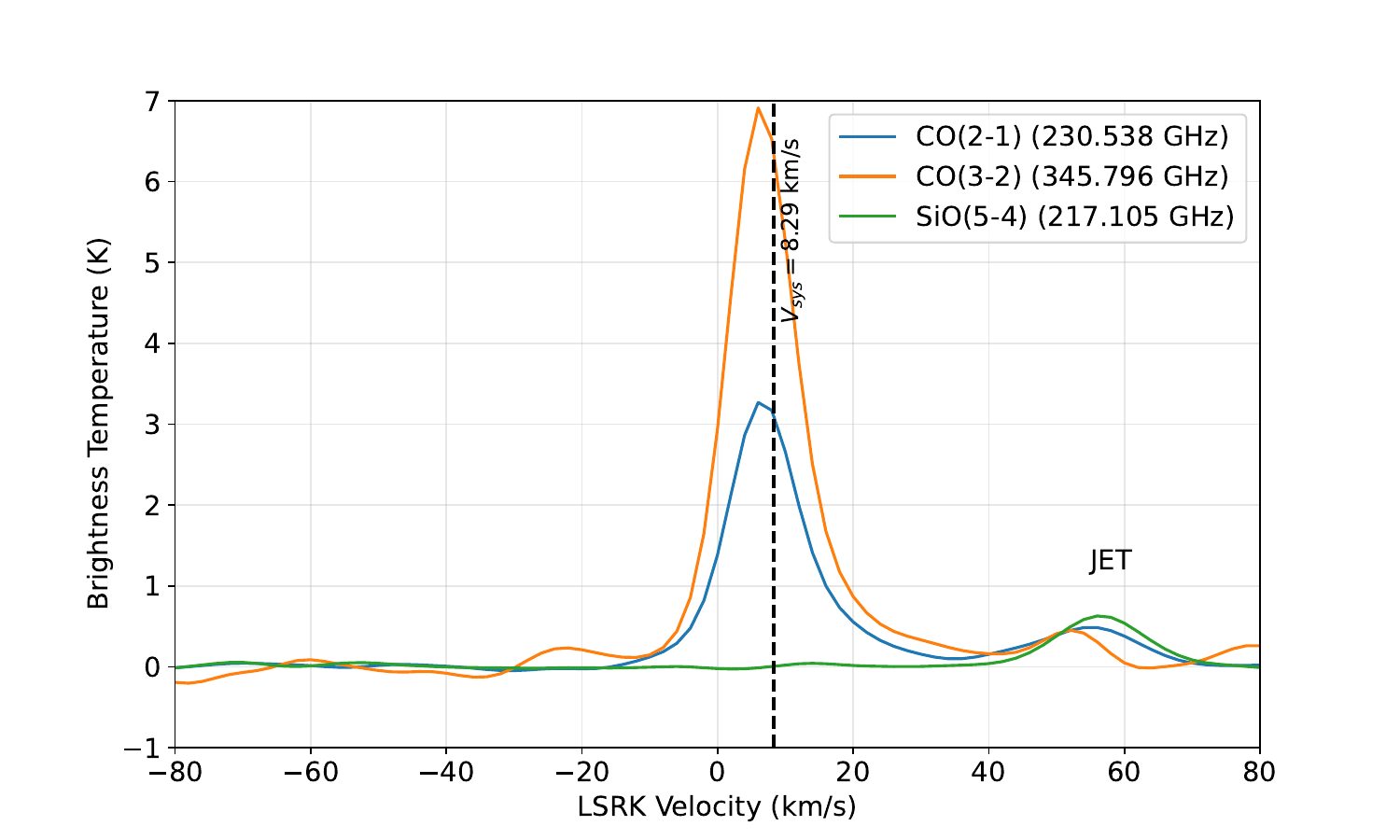}
\caption{Spectra of the CO(2--1), CO(3--2), and SiO(5--4) emission toward HOPS10 at $\sim$ 160 AU resolution, extracted from a $13\arcsec \times 1\arcsec$ region centered on the continuum peak and aligned with the jet axis (position angle $30^\circ$ east of north). All spectra are smoothed with a Gaussian kernel of $\sigma=2$ channels. The vertical dashed line marks the systemic velocity, $\rm V_{sys}=8.29~{km~s^{-1}}$.}
\label{fig:appendix_spectra}
\end{figure}

\section{Low-Resolution CO(2-1) Emission}\label{sec:appendix_ACACO_channelMap}
Figure~\ref{fig:appendix_ACACO_channelMap} shows the low-resolution ($\sim$2000~AU) $^{12}$CO(2–1) channel maps. The high-velocity jet emission is not detected toward the south–west  (the blueshifted lobe, south–west in the velocity range -100 to 0 km\,s$^{-1}$), while high-velcity (maps between +40 to +70 km\,s$^{-1}$) is clearly visible toward the north–east (the redshifted lobe, north–east in the velocity range 0 to +100 km\,s$^{-1}$). No extended cloud component is present in the direction of the blueshifted lobe that could be mixed up with the emission from the blueshifted side in the high-velcity range.

The additional CO emission seen toward the North or top-centre region (to the right of the redshifted outflow axis) in the low-velocity, $\rm V_{obs} = 0$ to $+10~{\rm km~s^{-1}}$ channel does not align with the jet/outflow axis and is spatially more diffuse than the collimated redshifted high-velocity jet or low-velocity outflow emission. We interpret this feature as likely arising from either the interaction of the low-velocity outflow cavity with ambient material or solely from ambient cloud material, rather than from the low-velocity outflow itself. There is no evidence for any extended or external dense cloud structure near the systemic velocity that could account for the absence of the blueshifted high-velocity lobe.

\begin{figure}[ht!]
    \centering
    \includegraphics[width=0.95\linewidth]{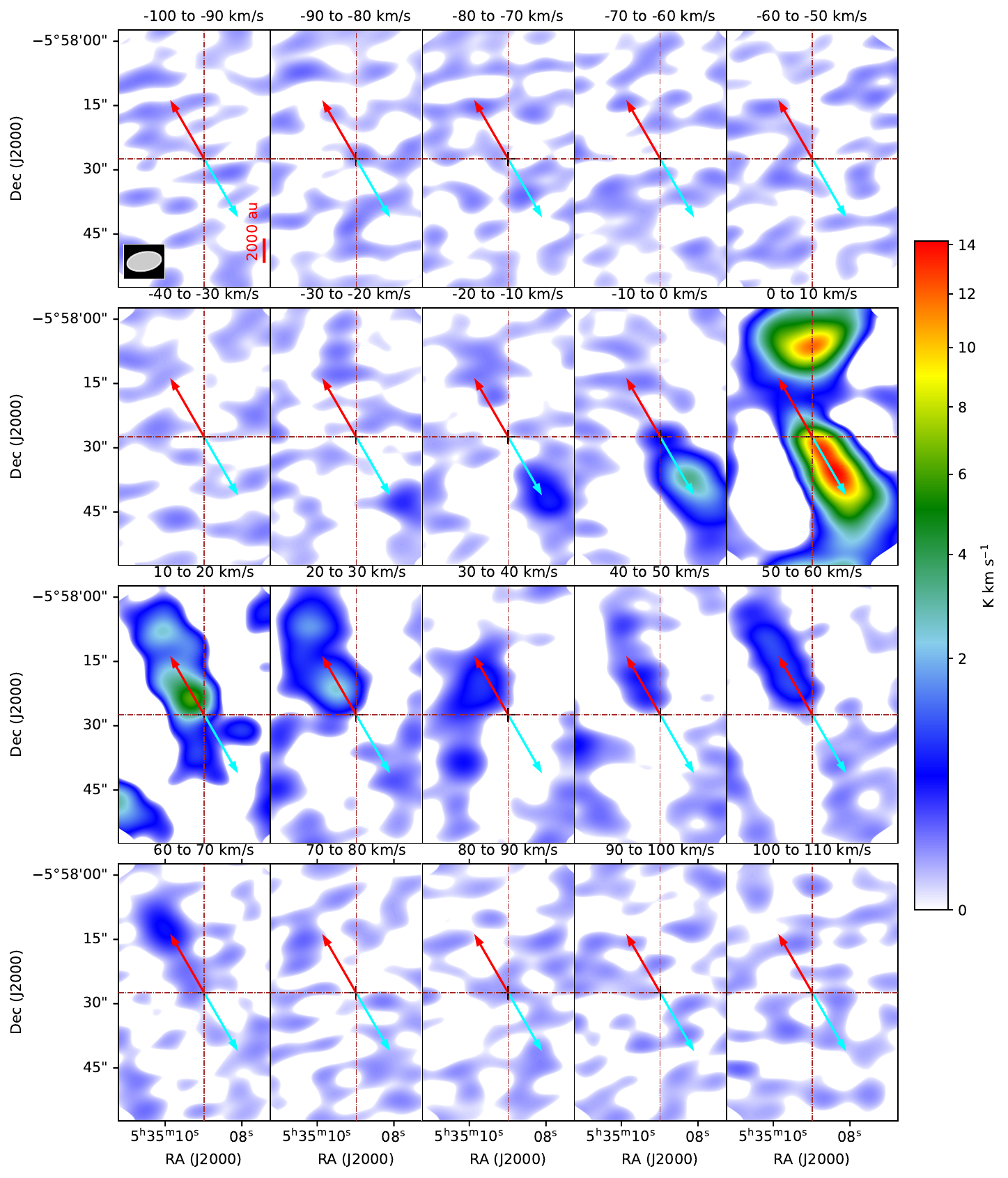}
\caption{ACA $^{12}$CO(2--1) channel maps at low resolution ($\sim$2000~AU).  The velocity range of each integrated channel, shown in the observed LSR (Local Standard of Rest) frame and not offset from the systemic velocity ($\rm V_{sys}=8.29~{km~s^{-1}}$), is indicated above the corresponding panel. The synthesized beam and the scalebar are shown in the first panel. Jet axis directions are indicated with red arrows for redshifted emission and cyan arrows for blueshifted emission. The continuum center is at the cross mark of the dotted lines, where the arrows originated.}
    \label{fig:appendix_ACACO_channelMap}
\end{figure}

\section{LTE Analysis of CO Lines}
\label{appendix_sec:opticaldepth}

To characterize the excitation and column density structure of the molecular outflow, we analyzed the CO~($J{=}3\!\rightarrow\!2$) and CO~($J{=}2\!\rightarrow\!1$) transitions under the assumption of Local Thermodynamic Equilibrium (LTE).  
The moment--0 maps of both transitions were beam-matched by convolving to the larger synthesized beam, and the CO~(3--2) map was reprojected onto the spatial grid of the CO~(2--1) map.  
Pixels below a $3\sigma$ threshold (set by the rms of each map) were masked.  
From the resulting maps, the line--ratio distribution was computed as
\begin{equation}
    R \equiv \frac{W_{32}}{W_{21}},
\end{equation}
where $W_{ul}$ denotes the integrated intensity (K\,km\,s$^{-1}$) of the $u\rightarrow\ell$ transition.

\subsection{Excitation Temperature from the Line Ratio}

Under optically thin LTE conditions, the intensity ratio of two CO rotational transitions is
\begin{equation}
    R
      = \frac{A_{32}\,\nu_{32}^{\,2}\,g_3}
             {A_{21}\,\nu_{21}^{\,2}\,g_2}
        \exp\!\left[-\frac{\Delta E}{T_{\rm ex}}\right],
\end{equation}
where $g_J = 2J+1$ and $\Delta E = E_{u,32} - E_{u,21}$ is the difference in upper--level energies.  
For $^{12}$CO, the Einstein $A$--coefficients are
\begin{equation}
    A_{21} = 6.91 \times 10^{-7}\ \mathrm{s^{-1}}, 
    \qquad
    A_{32} = 2.497 \times 10^{-6}\ \mathrm{s^{-1}},
\end{equation}
giving a constant ratio
\begin{equation}
    C \equiv \frac{A_{32}\,\nu_{32}^{\,2}\,g_3}
                 {A_{21}\,\nu_{21}^{\,2}\,g_2}
        \approx 9.0 .
\end{equation}
Solving the ratio relation for the excitation temperature yields
\begin{equation}
    T_{\rm ex}
      = \frac{\Delta E}{\ln(C/R)} .
\end{equation}

\subsection{Optically Thin LTE Column Density}

In the optically thin limit, the upper--level column density for a transition of frequency $\nu$ and Einstein coefficient $A_{ul}$ is
\begin{equation}
    N_u^{\rm thin}
      = \frac{8\pi k \nu^2}{h c^3 A_{ul}}\, W ,
\end{equation}
with integrated intensity $W$ converted to K\,m\,s$^{-1}$.  
The total CO column density follows from the Boltzmann distribution:
\begin{equation}
    N_{\rm tot}^{\rm thin}
      = N_u^{\rm thin}
        \frac{Q(T_{\rm ex})}{g_u}
        \exp\!\left(\frac{E_u}{T_{\rm ex}}\right),
\end{equation}
where $Q(T_{\rm ex})$ is the rotational partition function, defined as the total number of thermally accessible energy states (here, rotational) available to a molecule at a given temperature. 
For a linear rotor,
\begin{equation}
    Q(T_{\rm ex}) \approx \frac{T_{\rm ex}}{hB/k},
\end{equation}
with $B$ the rotational constant of CO.
\subsection{Peak Temperature and Optical Depth}

The peak brightness temperature can be estimated from the integrated intensity and linewidth:
\begin{equation}
    T_{\rm peak} \approx \frac{W}{\Delta v},
\end{equation}
where $\Delta v$ is the FWHM or effective linewidth.

The Planck brightness temperature is defined as
\begin{equation}
    J_\nu(T)
      = \frac{h\nu/k}{\exp(h\nu/kT)-1}.
\end{equation}

For a uniform LTE slab, the observed peak temperature is
\begin{equation}
    T_{\rm peak}
      = \bigl[J_\nu(T_{\rm ex}) - J_\nu(T_{\rm bg})\bigr]
        \left(1 - e^{-\tau_0}\right),
\end{equation}
where $T_{\rm bg}=2.73\,$K.

Inverting this equation gives the optical depth:
\begin{equation}
    \tau
      = -\ln\!\left[
        1 - \frac{T_{\rm peak}}
                 {J_\nu(T_{\rm ex}) - J_\nu(T_{\rm bg})}
        \right].
\end{equation}
This inversion is valid as long as
\begin{equation}
    0 \le x < 1, \quad
    x \equiv 
      \frac{T_{\rm peak}}
           {J_\nu(T_{\rm ex}) - J_\nu(T_{\rm bg})} = 1- e^\tau,
\end{equation}
because $x \ge 1$ corresponds to unphysical brightness temperatures exceeding the maximum allowed by $T_{\rm ex}$.

For the optically-thin limit, $\tau \ll 1$, the optical depth can be approximated by
\begin{equation}
    \tau_0 \approx
      \frac{T_{\rm peak}}
           {J_\nu(T_{\rm ex}) - J_\nu(T_{\rm bg})}.
\end{equation}
In summary,  the excitation temperature $T_{\rm ex}$ using Equ. C5 and the initial column density $\rm N_{tot}^{\rm thin}$  using Equ. C7 are derived under the optically thin LTE approximation.  The optical depth $\tau$ is computed independently from the measured $T_{\rm peak}$ using the exact radiative--transfer solution.




\bibliography{sample701}{}
\bibliographystyle{aasjournalv7}



\end{document}